\documentclass[12pt]{article}
\usepackage{epsfig}
\begin{document}
%set bwid 6; set fwid 6; set hwid 6; set pwid 6; set xwid 8; set ywid 6;
%igset lwid 6

\begin {center}
{\bf How Resonances can synchronise with Thresholds}
\vskip 5mm
{D.\ V.\ Bugg\footnote{email: d.bugg@rl.ac.uk} \\
{\normalsize\it Queen Mary, University of London, London E1\,4NS, UK}
\\[3mm]}
\end {center}
\date{\today}

\begin{abstract}
\noindent
The mechanism by which a threshold may capture a resonance is
examined.
It involves a threshold cusp interfering constructively with either
or both (i) a resonance produced via confinement,
(ii) attractive $t$- and $u$-channel exchanges.
The $f_0(980)$, $X(3872)$ and $Z(4430)$ are studied in detail.
The $f_0(980)$ provides a valuable model of the locking
mechanism.
The $X(3872)$ is too narrow to be fitted by a cusp, and
requires either a resonance or virtual state.
The $Z(4430)$ can be fitted as a resonance but also can be fitted
successfully by a cusp with no nearby resonant pole.

\vskip 2mm

\end{abstract}

\section {Introduction}
It is well known that several meson resonances appear at or close to
thresholds.
Examples are $f_0(980)$ and $a_0(980)$ at the $KK$ threshold,
$f_2(1565)$ at the $\omega \omega$ threshold and $K_0(1430)$ close to
the $K\eta'$ threshold.
In all these cases, decays to mesons are via S-waves.
Further possible examples are $X(3872)$ at the $\bar D(1865)D^*(2007)$
threshold, $Z(4430)$ close to the $D^*(2007)\bar D_1(2420)$ threshold
and $Y(4260)$ close to the $D(1865)\bar D_1(2420)$ threshold.
[The $\bar D D$ charge conjugate combination is tacitly included
throughout this paper].
Amongst baryons, $P_{11}(1710)$ and $P_{13}(1720)$ appear close to the
$N\omega$ threshold (1720 MeV), and $\Lambda_C(2940)$ appears close
to the $D^*(2007)N$ threshold.

All these cases have been discussed as `molecules', bound a few MeV
below the thresholds.
There is much argument whether they should be viewed as bound states
of mesons or as quark configurations, e.g. $c\bar c n\bar n$ in the
case of $X(3872)$.

The first objective of this paper is to draw attention to a
known but frequently overlooked mechanism which attracts resonances
to thresholds.
The $f_0(980)$ is studied in depth as one specimen.
The second objective is to point out the role of zero-point energy.
This provides some distinction between mesonic molecules (uncoloured)
and quark configurations, which are coloured and therefore confined.
Thirdly, an improved but convenient form of the Flatt\' e formula
\cite {Flatte} is proposed for the case of sharp thresholds like $KK$
and other cases cited above.

Consider as an example $f_0(980)$ and its decay to $KK$.
The conventional Flatt\' e denominator for this resonances is
%Eq. 1
\begin {equation}
D(s) = M^2 - s - i\sum _i \Pi _i ,
\end {equation}
where $M$ is resonance mass, $\rm {Im}\,\Pi_i = g_i^2\rho_i$, $g^2_i$
are coupling constants to decay channels and $\rho_i$ are phase space
factors.
However, further dispersive terms ${\rm Re}\, \Pi _i(s)$ are required
in $D(s)$:
%Eq 2
\begin {equation} -{\rm Re }\, \Pi _i(s) =
-\frac {1}{\pi} \rm {P} \int _{4m^2_i}
^\infty ds' \, \frac {g^2_i(s') \rho _i(s')} {s' - s},
\end {equation}
where $m_i$ are $\pi$ and $K$ masses and  P denotes the principal
value integral.
In fact, the terms $ig^2_i\rho_i$ of Eq. (1) arise from the pole at
$s' = s$ in Eq. (2).

Fig. 1 shows $\Pi _{KK}(s)$ and $g^2_{KK}\rho _{KK}(s)$ near the $KK$
threshold, using a form factor $e^{-3k^2}$, where $k$ is
centre of mass $KK$ momentum in GeV/c.
There is a cusp in $\Pi (s)_{KK}$ at the threshold.
The fact that it is positive definite at theshold signifies additional
attraction appearing there.
A minor technicality is that $\Pi (s)$ goes  negative after the
peak in $\rho(s)$, since  the dispersion integral is close to the
gradient of $\rho(s)$.
Eventually it returns slowly to zero for large $s$.
%Fig. 1
\begin{figure}[htb]
\begin{center} \vskip -12mm
\epsfig{file=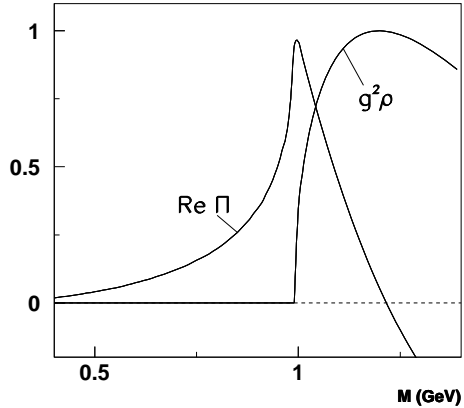,width=8cm}
\vskip -6mm
\caption {${\rm Re}\,  \Pi _{KK}(s)$ and $g^2_{KK}\rho _{KK}(s)$
for $f_0(980)$, normalised to 1 at the peak of $g^2_{KK}\rho _{KK}$.}
\end{center}
\end{figure}

If the cusp is superimposed on attraction from another
source, for example meson exchanges, a resonance can be generated by
the peak in $\rm {Re}\, \Pi$; if the attraction is not quite
sufficient to produce a resonance, there is a virtual state instead.
The $\sigma$, $\kappa$, $a_0(980)$ and $f_0(980)$ make good
candidates for such molecules.
Otherwise, all known cases except $Z(4430)$ can conservatively be
accomodated as regular quark resonances whose masses have been
perturbed to synchronise with thresholds.

Near threshold, the wave function of the resonance has a long tail,
like the deuteron.
The long-range tail is purely mesonic; any contribution to the wave
function from coloured quarks is confined in the resonance
at short range or in the decay mesons.
In the confined part of the wave function, quarks have kinetic
energy $k^2/2m = -(\hbar ^2/2m)\nabla ^2 \Psi$; here $k$ and $m$ are
quark momentum and effective mass and $\Psi$ is the wave function.
If a bound state is to be created, extra potential energy is needed to
compensate the kinetic energy of the confined particles.
This again pulls the resonance towards the threshold.

Jaffe proposed that the $f_0(980)$ and $a_0(980)$ are
members of a 4-quark nonet with composition $s\bar s(u\bar u \pm
d\bar d)$ \cite {Jaffe}.
This may be correct, but a long-range tail of $KK$ is unavoidable.

Section 2 reviews the formalism.
This leans heavily on a perceptive paper of T\" ornqvist
\cite {Tornqvist}, written in 1995, before many of the examples
cited above were known or accurately measured.
For the case of $Z(4430)$, some changes need to be introduced
into his formalism.

Section 3 examines the $f_0(980)$ in detail.
Perturbations around its fitted parameters illustrate
the general way in which a threshold captures a resonance.
Section 4 examines $X(3872)$, which peaks within 0.6 MeV of the
$\bar D D^*$ threshold.
It could be a mesonic molecule, as T\" ornqvist \cite {Torn2},
Close and Page \cite {Close} and Swanson \cite {Swanson}
have argued.
Its  very close relation to the threshold points strongly to the
cusp mechanism playing a decisive role.
A likely possibility is that the $\bar cc ~ ^3P_1$ radial
excitation has been captured by the $\bar D D^*$ threshold.
If that state were to appear elsewhere, the alternative explanation
is a mesonic molecule or virtual state.
Present data simply require a second-sheet pole.

The $Z^+(4430)$ reported by Belle \cite {Belle} appears as a peak in
the $\Psi '\pi ^\pm$ mass spectrum in $B \to \Psi' \pi ^\pm K$.
The remarkable feature of this peak is that it has isospin 1,
unlike regular $c\bar c$ states, so it is exotic.
Belle fit it as a resonance with mass $M = 4433 \pm 4 \pm 1$
MeV and width $\Gamma = 44 ^{+17 \, +30} _{-13 \, -11}$
MeV.
Rosner \cite {Rosner} and also Meng and Chao \cite {Meng}
point out that its mass is close to that of $D^*(2010) +
\bar D_1(2420)$.
Maiani et al. \cite {Maiani} interpret the peak as a
diquark-antidiquark state.
For an S-wave $D^*\bar D_1$ combination, its spin-parity $J^P$
is $0^-$, $1^-$ or $2^-$.
It is discussed in detail in Section 5.
It is easily fitted as a resonance.
However, it can also be fitted as a cusp due to strong de-excitation
of $D^*(2010)\bar D_1(2420)$ to lower configurations of
$D\bar D$ and $D^*\bar D$.

Section 6 makes remarks about a number of the other states associated
with thresholds, so as to fill in the broader picture.
Section 7 summarises conclusions.

\section {How to parametrise a cusp or threshold resonance}
T\" ornqvist's formalism \cite {Tornqvist} will be followed here
with some modifications.
He considers the possibility of overlapping resonances
such as $f_0(980)$ and $f_0(1370)$;
that is a detail not required here, simplifying the notation.
His formula for the amplitude connecting channels $i$ and $j$ (e.g.
$\pi \pi$ and $KK$) is then
%Eq 3
\begin {equation}
A _{ij}(s) = T_{ij}(s)\sqrt {\rho_i \rho_j} = G_i^\dagger (s) P G_j(s),
\end {equation}
where $P$ is the propagator; $T$ is the usual $T$-matrix with phase
space factored out.
The $G_i$ are related to coupling constants $g_i$ via
%Eq 4
\begin {equation}
G^2_i (s) = g^2_i \rho_i (s)F^2_i(s)\theta (s - s_{th,i}),
\end {equation}
with $s_{th,i}$ the threshold for channel $i$; $F_i(s)$ is a form
factor, taken here as $\exp (-k^2R^2/6)$ with $R = 0.6$ fm.
Then $F^2 \simeq \exp(-3k^2)$ with $k$ in GeV/c.
For narrow resonances, the precise $s$-dependence
of $F^2$ is not decisive but begins to matter for
broader structures like $Z(4430)$.

For a resonance,
%Eq 5
\begin {equation}
P^{-1}(s) = M^2 - s - \Pi (s).
\end {equation}
The definition of $\Pi$ has the opposite sign to that used by
T\" ornqvist, to avoid a multiplicity of minus signs in formulae
which follow.
The unitarity relation ${\rm Im}\, A = A A^\dagger$ gives
%Eq 6
\begin {equation}
\rm {Im}\, \Pi(s)  = \sum _i g^2_i \rho_i(s) F^2_i(s)\theta (s -
s_{th,i}).
\end {equation}
Using analyticity,
%Eq 7
\begin {equation}
\rm {Re}\,  \Pi(s)  = \frac {1}{\pi} {\rm P} \int _{s_{th,i}}^\infty
ds' \frac {\rm {Im}\, \Pi (s')}{s' - s}.
\end {equation}
The term inside the summation of Eq. (6) is positive
definite , leading always to attraction at and below
threshold,
though the dispersion integral can change sign above threshold.

\subsection{Useful resonance formulae}
If $F^2(s) = 1$, the integral in (7) diverges.
However, with two subtractions, it can be evaluated analytically
with the result \cite {Cabibbo}
%Eqs 8 and 9
\begin {eqnarray}
\frac {\rm {Re}\, \Pi(s)_i}{g^2_i}  = j_i &=&
\frac {\rho_i}{\pi} \ln \frac
{1-\rho_i}{1 + \rho _i }, \qquad s \ge s_{th,i} \\
&=& -\sqrt {\frac {4m^2_i -s}{s} } 
-\frac {2v_i}{\pi} \tan ^{-1} v_i, \qquad s < s_{th,i}
\end{eqnarray}
where
%Eq 10 and 11
\begin {eqnarray}
\rho _i &=& 2k_i/\sqrt {s}, \qquad s \ge s_{th,i} \\
v_i &=& 2|k_i|/\sqrt {s}, \qquad s < s_{th,i}
\end{eqnarray}
and $k_i$ are momenta of decay particles in the rest frame of the
resonance.
The first term of Eq. (9) is the usual Flatt\' e extrapolation 
below threshold.
The contributions to $j_i$ near threshold from remaining terms are
%Eq 12 and 13
\begin {eqnarray}
\pi j _i &=& -2\rho^2_i - (2/3)\rho ^4_i + \ldots, \qquad s \ge s_{th,i}
\\
 &=& -2v^2_i + (2/3)v^4_i + \ldots, \qquad s < s_{th,i} .
\end{eqnarray}
The first term in each is symmetrical about threshold, but the
next term leads to an asymmetry away from the threshold.
It is also important to realise that the two subtractions used in
arriving at this result allow $\Pi _i$ to contain an additional
linear dependence on $s$.
A constant term is certainly required to make the
integral of (7) positive at threshold.

In illustrations given later, the principal value integral will be
evaluated numerically.
However, the algebraic forms (8) and (9) improve on the usual
Flatt\' e form for a resonance near threshold, but are equally
convenient to use.
A constant term can be absorbed into $M^2$ of the Breit-Wigner
amplitude giving
%Eq 14
\begin {equation}
D(s) = M^2 - s - \sum _i g^2_i(j_i + i\rho _i).
\end{equation}
For small $v_2^2$ below the opening threshold, the second term of
Eq. (9) contributes
%eq. 15
\begin {equation}
\frac {2}{\pi}g_2^2v^2_2 = \frac {2g_2^2}{\pi}\left(
\frac {4m^2_K -s}{s}\right).
\end {equation}
Note the explicit appearance of the factor $(4m^2_K - s)$ in
equation (15) and also in the first term of Eq. (9).
The cusp mechanism is contributing to the resonance through a
term like $(M^2 - s)$ but peaking at the threshold.
Also note that, when fitting data, there is a strong correlation
between the term $M^2 - s$ and $-g_2^2j_2$.
If there are no data on the $KK$ channel to determine $g_2^2$ directly,
data below the $KK$ threshold give almost no determination of $g_2^2$
in view of the poorly known form factor $F^2_2$.

The fit to BES data on $J/\Psi \to \phi \pi ^+\pi ^-$ and
$\phi K^+K^-$ \cite {phipp} changes very little when the Flatt\' e
formula is replaced by Eq. (9).
The $\phi KK$ data deterine $g^2_2$ well, and a modest change in $M^2$
compensates for the change in the formula.

If there is a linear subtraction term in addition to the constant term,
%Eqs 16 and 17
\begin {eqnarray}
D(s) &\to& M^2 - (1 + \beta)s - \sum _i g^2_i(j_i + i\rho _i) \\
 &=& (1 + \beta )\left[\frac {M^2}{1 + \beta} - s
 - \frac {1}{1 + \beta }\sum _i g^2_i(j_i + i\rho _i)\right].
\end{eqnarray}
This leads to a renormalisation of the mass and width, as observed
for $f_0(1370)$ in a recent re-analysis including dispersive terms
arising from the opening of the $4\pi$ channel \cite {f01370}.
For practical purposes, the factor $1/(1 + \beta )$ renormalises
mass and width terms in (17).

\subsection {No resonance and approach to resonance}
If there is no resonance, it is convenient to replace $M^2 - s$
of (14) by a constant $M^2$, hence keeping dimensions unchanged.
In Section 5, it will be shown that the K-matrix approach leads to
further terms neglected by T\" ornqvist, but they can safely be
ignored in the cases of $f_0(980)$ and $X(3872)$.
Standard effective range theory replaces
$M^2$ by $M^2 - \gamma k^2$.
As $\gamma$ increases, a virtual state approaches the threshold
and smoothly becomes a bound state - or resonance if
there are open channels.

The cusp mechanism gives maximum attraction at the threshold.
A resonance close to threshold necessarily has a long-range tail,
mesonic in character.
Consider as an example mixing between $c\bar c$ and a
$(c\bar n)(\bar c n)$ configuration, appropriate to the discussion of
$X(3872)$.
The mixing is given by the familiar eigenvalue equation
%Eq 18
\begin {equation}
H \Psi  = \left(
\begin {array} {cc}
H_{11} & V \\
V & H_{22}
\end {array}
\right) \Psi,
\end{equation}
where $H_{11}$ and $H_{22}$ describe isolated $\bar cc$ and
$(c\bar n)(\bar c n)$ configurations and $V$ describes mixing.
This mixing pushes the $c\bar c$ state down in mass.
This is the well known Variational Principle which minimises the
eigenvalue when states mix.
It is closely analogous to formation of a covalent bond in chemistry.
The $c\bar n$ configuration is different in detail from an atom made
of a proton and electron, but the principle is the same.
The detailed dynamics of the decay of a confined state through
the confining barrier is a key missing detail at present; so is the
possible role of di-quark interactions in the `molecule'.

The meson-meson component has $t$ or $u$ channel poles, but slow
variation with $s$.
The linear combination with $c\bar c$ has minimum zero-point energy
when the resonance is centred on the threshold {\it and} is narrow.
Both the cusp attraction at threshold and the effect of zero-point
energy provide feedback tending to lock the resonance to the
threshold.
In the next section, the $f_0(980)$ will be used as a model to
examine this locking mechanism numerically.

The dispersive effect at the threshold may be an unfamiliar effect.
Vacuum polarisation arises in this way.
An example from classical physics is a tsunami - a travelling cusp.
As a tsunami approaches a beach, attraction into the wave-front
drains water from the beach well in advance of the wave and gives
advance warning of the approaching wave.

\section {The $f_0(980)$}
Can we be sure that the $f_0(980)$ is really a resonance and not
a virtual state or cusp?
Cern-Munich data on $\pi \pi$ elastic scattering \cite {Hyams}
were first fitted to a simple Breit-Wigner resonance of constant width
and yielded a good fit with a phase increase of $\sim 180^\circ$
over the $KK$ threshold.
Crystal Barrel data on $\bar pp \to 3\pi  ^0$ are a further delicate
source of information \cite {Abele}.
The  $f_0(980)$ appears as a visible dip in the $\pi \pi$ mass
projection.
These data provide accurate phase information on $f_0(980)$ via
interferences between the three $\pi^0 \pi ^0$ combinations all over
the Dalitz plot.
%Fig. 2
\begin{figure}[htb]
\begin{center}
\vskip -10mm
\epsfig{file=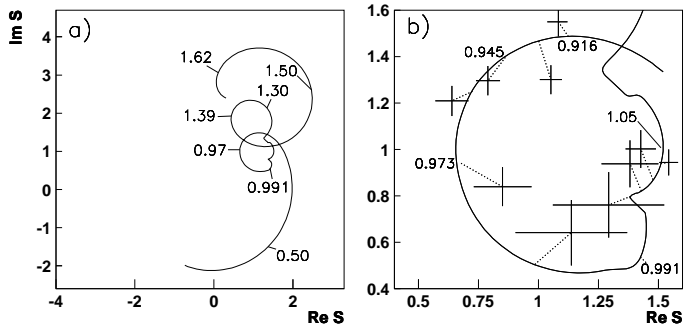,width=10cm}
\vskip -6mm
\caption {(a) The Argand diagram for the $\pi \pi$ S-wave from
[13], (b) enlarged comparison with the $f_0(980)$ mass
region; crosses indicate errors for free fits to real and imaginary
parts of the amplitude in 10 equal bins of $s$ from 0.84 to 1.08 GeV$^2$;
dotted lines indicate the movement of free fits from that
with a Flatt\'e formula;
numbers indicate masses in GeV.}
\end{center}
\end{figure}

Fig. 2(a) shows the Argand plot of the $\pi \pi$
S-wave amplitude from a recent re-analysis of the data \cite {f01370}
using a Flatt\' e formula for $f_0(980)$.
As a test of the resonance hypothesis, these data have been refitted
allowing complete freedom for the real and imaginary parts of the
$f_0(980)$ amplitude in 10 equal bins of $s_{\pi \pi}$ from
0.84 to 1.08 GeV$^2$.
Fig. 2(b) compares results (crosses indicating the errors) with the
fit of \cite {f01370} shown by the smooth curve.
The errors allow for a mass resolution of $\pm 4$ MeV, which has
quite a large effect near the $KK$ threshold.

It is clear from Fig. 2(b) that the phase shift of $f_0(980)$
increases by $\sim 90^\circ$ from $\sim 900$ MeV to the $KK$
threshold at 991 MeV (the average mass for $K^+K^-$ and
$K^0 \bar K^0$).
This supports the resonance interpretation.

As a further test, the fit has been repeated using Eqs. (8) and
(9), which improve on the Flatt\' e formula.
Also, separate thresholds have been included for $K^+K^-$ and
$K^0 \bar K^0$.
There is a small improvement of 23 in log likelihood (statistically
5.5 standard deviations).
There is only a small shift in the second sheet pole, which moves
from $998.4 \pm 4 - i(17.2 \pm 4)$ MeV to $(1003.9 - i16.5)$ MeV
with identical errors.
[For the second sheet pole, the signs of $\rho _2$ and $j_2$
are reversed;
for the third sheet pole, they are as in Eq. (14)].
Because the parametrisation of the cusp is now included,
the distant third sheet pole moves from $(851 - i418)$
MeV to $(1171 - i592)$ MeV; this is because the cusp formula modifies
the shape of the resonance away from the $KK$ threshold.

%Table 1
\begin{table}[htb]
\begin {center}
\begin{tabular}{ccc}
M & $g^2_2$ & Pole \\
(GeV) &  (GeV$^2$) & (MeV) \\
\hline
0.956 &  0.895 & $778 - i42$ \\
      &  1.095 & $825 - i36$ \\
      &  1.295 & $858 - i31$ \\
      &  1.595 & $892 - i25$ \\
\hline
      &  0.695 & $710 - i48$ \\
      &  0.495 & $605 - i52$ \\
      &  0.295 & $438 - i44$ \\
      &  0.195 & $318 - i24$ \\
\hline
\end{tabular}
\caption{Movement of the second sheet pole of $f_0(980)$ with $g^2_2$
using a pure cusp formula.}
\end {center}
\end{table}

It is of interest to use the $f_0(980)$ as a model to examine
the movement of the second-sheet pole as the parameters of the
resonance are varied. This reveals the general features of the way
the resonance forms.
These general features are likely to be similar for other cases.

A fundamental point is whether the cusp alone can produce a pole at
or very close to the $KK$ threshold.
The answer is no.
For this test, the term $-s$ of the Breit-Wigner denominator,
Eq. (14), is removed, again including $KK$ mass differences.
A second sheet pole does appear, but well below the $KK$ threshold.
Table 1 shows the pole position for a variety of values of $g^2_2$.
For small $g^2_2$, the pole moves rapidly away
from the $KK$ threshold.
The reason the pole always lies below the $KK$ threshold is that it
is closely related to the term $2|k_2|/\sqrt {s}$ of Eq. (11)
which increases below the $KK$ threshold with $|k_2|$.

The conclusion is that formation of the resonance requires an
additional source of attraction, e.g. meson exchanges.
Janssen et al. \cite {Janssen} were able to generate a
resonance from $K^*$ and $\rho$ exchanges between $\pi \pi$ and $KK$.
Likewise, the $\sigma$ and $\kappa$ poles appear from meson exchanges
in the calculations of Caprini et al. \cite {Caprini}
and B\"uttiker et al. \cite {Buttiker} using the Roy equations.
%Table 2
\begin{table}[htb]
\begin {center}
\begin{tabular}{ccc}
M & $g^2_1$ & Pole \\
(MeV) & (GeV$^2)$ & (MeV) \\
\hline
0.50 & 0.185 & $806 - i76$ \\
0.60&        & $852 - i68$ \\
0.70  &      & $899 - i59$ \\
0.80  &      & $946 - i48$ \\
0.85  &      & $968 - i41$ \\
0.90  &      & $987 - i31$ \\
0.94  &      & $1000 - i21$ \\
0.956 &      & $1004 - i17$ \\
0.97  &      & $1007 - i12$ \\
0.99  &      & $1011 - i4$ \\
1.01  &      & $1012 - i4$ \\
1.03  &      & $1012 - i15$ \\
1.05  &      & $1009 - i28$ \\
1.10  &      & $979 - i69$ \\
\hline
0.956 & 0.285& $1023 - i32$ \\
      & 0.1  & $993 - i7$ \\
      & 0.05 & $989 - i3$ \\
      & 0.02 & $988 - i0.8$ \\
      & 0    & 987.2        \\
\hline
\end{tabular}
\caption{Movement of the second sheet pole of $f_0(980)$
from Eqs. (8) and (9) as its parameters are varied. In all cases,
$g^2_2 = 0.875$ GeV$^2$.}
\end {center}
\end{table}

The attraction due to these long-range forces makes the real part of
$D(s)$ pass through 0 at some mass.
It is of interest to see how effective the cusp is in attracting
the second sheet pole when this crossing point varies.
Table 2 shows the pole position as $M$ of the Breit-Wigner formula
is varied.
From the first few entries, one sees that the pole moves a long way
if $M$ is far from the $KK$ threshold.
The cusp is effective in attracting the resonance over a
mass range much larger than the $\pi \pi$ width.
From $M=0.9$ to 1.10 GeV, the pole stays quite close to the $KK$
threshold.
For $M$ outside the range 0.5 to 1.1 GeV, it disappears.
The pole is more easily pulled up to the $KK$ threshold than down to
it.
Over the range of values in the Table, the distant third sheet pole
moves only by small amounts.

The bottom entries in the Table show how the second sheet pole varies
with $g^2_1$.
As it decreases, the width of the pole decreases more rapidly;
as $g_1^2 \to 0$, it becomes a bound state 0.15 MeV below the $K^+K^-$
threshold (the lower of the two $KK$ thresholds).
This bound state survives unchanged for zero $g^2_1$ for any value
of $M$ over the range 0.5 to 1.1 GeV and beyond.

The general pattern which emerges is that a cusp superimposed on
attraction due to either meson exchanges or quark confinement can
lock the resonance close to the threshold for a wide range of
parameters $M$ and $g^2_2$.
The cusp acts as a trigger for the resonance.
This is particularly the case when $g^2_1$ is small.
The locking mechanism is obviously inhibited if there is repulsion in
meson exchanges.

T\" ornqvist \cite {Tornqvist} gives a formula for the $KK$ component
in the wave function of any resonance close to a threshold.
For the present case, it may be written:
%Eq. 19
\begin {equation}
\psi = \frac {|q\bar q q \bar q> +\sum _i
[(d/ds)\rm {Re}\, \Pi_i(s)]^{1/2} |KK>} {1 + \sum_i (d/ds)\rm {Re}\,
\Pi_i(s)};
\end {equation}
this was evaluated in Ref. \cite {Decays}
and gives $\sim 60\%$ $KK$ component for $f_0(980)$.
That reference also shows that decay branching ratios favour an
abnormally large $KK$ component in $f_0(980)$.

Model calculations of a similar nature have been made by van Beveren
and Rupp \cite {eef}.
They adopt a transition potential coupling confined states in a
harmonic oscillator potential to outgoing waves, with a matching
at the transition radius $\sim 0.65$ fm.
The $\sigma$, $\kappa$, $a_0(980)$ and $f_0(980)$ emerge from the
continuum as their coupling constant to confined states is increased.
The great merit of their model is that it is straightforward to
follow the movement of poles as this coupling constant varies.
They give graphic illustrations of the movement of the $\kappa$ and
$a_o(980)$ poles with coupling constants.
The movement of the $\sigma$, $\kappa$,
$a_0(980)$ and $f_0(980)$ poles as a function of coupling constant is
also tabulated in Ref. \cite {joint}.
The model reproduces fairly well the amplitudes for all these states
with a universal coupling constant.
The $a_0$ does not appear at the $\eta\pi$ threshold because of the
nearby Adler zero.
The $\sigma$, $\kappa$ and $a_0$ all become bound states if
the coupling constant is increased by a factor 2.5.

Jaffe proposes that mesons can be divided into `ordinary mesons'
which decouple from scattering channels as the number of colours
$N_c \to \infty$ and `extrordinary mesons' which disappear in this
limit \cite {Jaffe2}.
He draws attention to the work of Pelaez on the $N_c$ dependence
of the chiral Lagrangian and its predictions for low mass
$\pi \pi$ scattering \cite {Pelaez}.

A final comment is that the phase shift for $\sigma$ goes through
$90^\circ$ very close to the $KK$ threshold.
This may not be accidental.
The $\sigma$ has significant coupling to $KK$ and it is possible
that the cusp mechanism is sufficient to tie the $90^\circ$ phase
to the threshold as well as that of $f_0(980)$.

\section {$X(3872)$}
Fig. 3 shows $j_2$ and $\rho _2$ for the $\bar D D^*$ threshold.
The result is similar to Fig. 1.
Attempts to fit the data with a bare cusp (i.e. with $M^2 -s$ of
the denominator replaced with just $M^2$) can reproduce the very
narrow width observed in decays to $\rho J/\Psi$ only with very fine
tuning of parameters.
There is then a second sheet pole almost at the $D\bar D^*$ threshold.
However this narrow width is incompatible with data on decays
to $\bar D D^*$ shown below in Fig. 4.
The conclusion is that a resonance or virtual state is required

The cusp can however capture a nearby $c \bar c$ $^3P_1$ state.
The alternative possibility is a molecule generated by $\pi$
exchange.
%Fig. 3
\begin{figure}[htb]
\begin{center}
\vskip -12mm
\epsfig{file=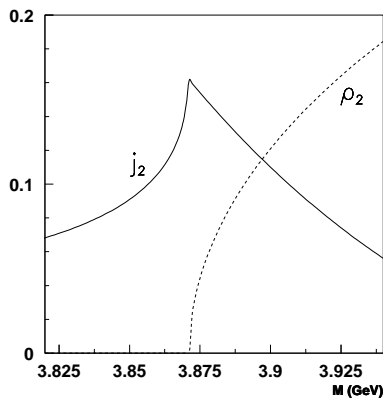,width=6cm}
\vskip -3mm
\caption {$j _{\bar D D^*}(s)$, $\rho _{\bar D D^*}(s)$ for $X(3872)$;
the vertical scale is arbitrary.}
\end{center}
\end{figure}
%Fig. 4
\begin{figure}[htb]
\begin{center}
\vskip -12mm
\epsfig{file=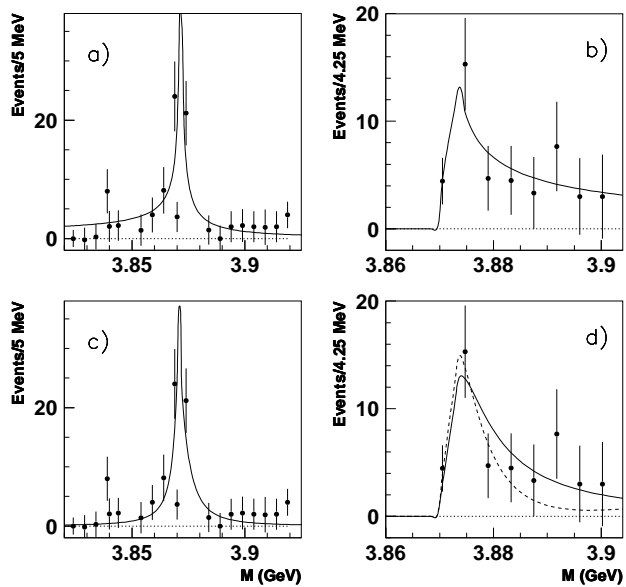,width=9cm}
\vskip -5mm
\caption {Fits to Belle data for (a) and (c)
$X(3872) \to \rho J/\Psi$,
(b) and (d) $\bar D D^*$; the upper panels are fitted with a virtual
state and the lower two with a resonance.}
\end{center}
\end{figure}

The data with the best mass resolution are from Belle and Babar on
decays to $\rho J/\Psi$ \cite {Choi} \cite {Aubert} and
$\bar D D^*$ \cite {Gokhroo} \cite {Grenier}.
These have been fitted by Braaten and collaborators \cite {Braaten}
and by Hanhart et al. \cite {Hanhart}.
They favour respectively a weakly bound state and a virtual state.
Both omit the dispersive effect discussed here, but assume an
attractive interaction which generates a nearby second sheet pole.

It is necessary to fit the narrow width observed in $\rho
J/\Psi$ simultaneously with the $\sim 3.5$ MeV mass difference between
the peaks observed there and in $\bar D D^*$.
The Belle data are shown in Fig. 4 with a variety
of fits.
For $\rho J/\Psi$ the improved data set shown by Olsen is used
\cite {Olsen}.

The mass distribution in $\rho J/\Psi$ (taken as channel 1)
and $\bar D D^*$ (channel 2) are:
%Eqs. 20-22
\begin {eqnarray}
\frac {dN _1}{ds} &=& \Lambda \frac {G_1^2 \rho _1(s)}{|D(s)|^2} \\
\frac {dN _2}{ds} &=& \Lambda \frac {G_2^2 \rho _2(s)}{|D(s)|^2} \\
D(s) &=& M^2 - s - i M(\Gamma _1 + g^2_2[\rho _i(s) + j_2/i])
\end {eqnarray}
for a resonance; for a virtual state, $-s$ in $D(s)$ is replaced by
$-\gamma k^2_2$.
The parameter $\Lambda$ is an overall normalisation constant.
A constant $\Gamma_1$ describes the width to channels other than
$\bar D D^*$.

The peak in $\bar D D^*$ is naturally higher in mass than in
$\rho J/\Psi$ because of the factor $\rho_2$ in the numerator of
Eq. (21).
However the width of the $J/\Psi$ peak cannot be too small if the
difference in peak positions is to be accomodated.
The branching ratio $BR[\rho J/\Psi]/BR[\bar D D^*] = 0.08 \pm 0.04$
including allowance for $D^* \to D\gamma$ as well as $D\pi$.
The branching ratio $BR[\omega J/\Psi ]/BR[\rho J/\Psi ] =
1.0 \pm 0.4 \pm 0.3$ \cite {Abe}.
A fit with these decay modes gives only a 1.5 MeV mass difference
between $\rho J/\Psi$ and $\bar D D^*$, but is sensitive to the
assumed mass resolution; here a moving average is calculated over the
4.25 MeV bins quoted by Belle for $\bar D D^*$ data.

There may well be other unobserved decays, e.g. to
$[\chi _1 \sigma]_{L=0}$ and/or $[\eta _C \sigma]_{L=1}$,
where $L$ is the orbital angular momentum in the decay;
they would be hard to detect experimentally.
Contributions from these channels seem likely since decays to
$\rho J/\Psi$ are isopin violating and the phase space
for $\omega J/\Psi$ is very limited.
The splitting between $\rho J/\Psi$ and $\bar D D^*$ peaks is easily
increased to 3 MeV by allowing extra decay channels with a rate a
factor 4-8 larger than $\rho J/\Psi + \omega J/\Psi$.

Figs. 5(a) and (b) show fits with a virtual state;
Figs. 5(c) and (d) use a resonance.
There is little to choose between them.
Presently the only way of making a definite distinction would be to
determine the phase directly from interferences with other components
in Dalitz plots; the virtual state has a falling phase above the peak.
A minor detail is that Fig. 4(d) show a second fit as a dashed curve,
assuming there is some background in $\bar D D^*$ at high masses.
In view of the uncertainties, a full list of fitting parameters is
not useful.
For the full curve of Fig. 4(d),
$M = 3871.8$ MeV, $\Gamma = 8$ MeV and equal branching ratios are used
to $\bar D D^*$ and the sum of other channels.
The full-width at half-maximum is however only 4.3 MeV because of
the cusp contribution to the resonance.

Belle report a peak in $\bar DD^*$ at $3942 ^{+7}_{-6} \pm 6$ MeV with
$\Gamma = 37 ^{+26}_{-18} \pm 8$ MeV \cite {Pakhlov}.
A spin-parity analysis is important.
It could be the $^3P_1$ radial excitation, in which case $X(3872)$ is
to be interpreted as a molecular state generated by the cusp mechanism
and attractive meson exchange.
Alternatives are $J^P = 0^-$, $1^-$ or $2^-$, though these would
be suppressed near the $\bar D D^*$ threshold by an $L=1$ centrifugal
barrier for decay.

\section {$Z(4430)$}
The $Z^+(4430)$ observed recently by Belle is important as a clear
candidate for an exotic meson.
It appears at the $\bar D^*(2010) D_1(2420)$ threshold within
errors.
It has a modest width $\sim 44$ MeV comparable with the width
of $D_1$ itself $(25 \pm 5)$ MeV.

Unlike $X(3872)$ it has many likely de-excitation processes
\begin {itemize}
{\item $0^- \to [D\bar D_0(2308)]_{L=0}$, $[D\bar D^*]_{L=1}$ and
$[D^*\bar D^*]_{L=1}$,} \\
{\item $1^- \to [D^*\bar D_0(2308)]_{L=0}$,
$[D\bar D]_{L=1}$, $[D\bar D^*]_{L=1}$ and $[D^*\bar D^*]_{L=1}$,} \\
{\item $2^- \to [D\bar D^*]_{L=1}$ and $[D^*\bar D^*]_{L=1}$.} \end
{itemize}
The essential point is that these de-excitation processes unavoidably
lead to a strong threshold cusp unless meson exchanges are repulsive
for all $J^P$.

It is necessary to fold the width of the $D_1(2420)$ into the
calculation of the cusp.
Consider as an example decays of $D^*(2010)D_1(2420)$ to $D^*(2010)
D(1865)$ with $L=1$.
For fixed total $s$ of the initial state and fixed mass $s_1$ of $D_1$,
the intensity is proportional to $(2k_1/\sqrt {s})B_1(k_1R)F^2(k_1R)$
(ignoring the very small widths of $D $ and $D^*$);
here $k_1$ is the centre of mass momentum in the final state
and $B_1 = k_1^2R^2/(1 + k_1^2R^2)$ is the centrifugal barrier factor
for $L=1$ decay.
Since $k_1$ is large ($\sim 1070$ MeV/c), $B_1 \simeq 1$.
Including the phase space for the initial state and integrating over
the line-shape of $D_1(2420)$, the phase space factor for the
whole process is
%Eq. 23
\begin {equation}
\rho (s) = \int _{(M_D +M_{D^*})^2} ^{(\sqrt {s} - \sqrt {s_1})^2}
ds_1 \frac {F^2(k_2R)}{|D(s_1)|^2} \frac {4 k_2 k_1}{\sqrt {s s_1}}
F^2(k_1R),
\end {equation}
where $k_2$ is the momentum in the $D^*\bar D_1$ channel.
This integral is easily done numerically.
The important dependence is on $k_2$, $F^2(k_2R)$ and $D(s_1)$.

%Fig. 5
\begin{figure}[t]
\begin{center}
\vskip -12mm
\epsfig{file=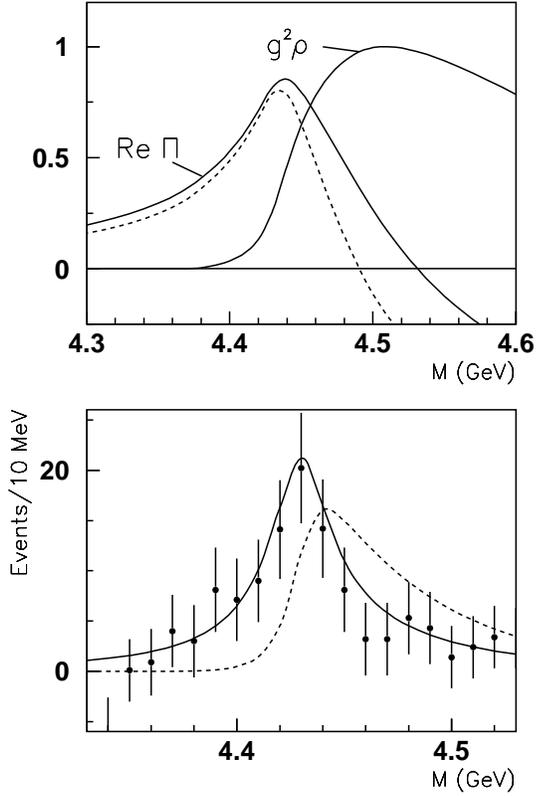,width=8cm}
\vskip -7mm
\caption {(a) $g^2(D^*\bar D_1)$ and ${\rm Re}\, \Pi (D^*\bar D_1)$
calculated from de-excitation of $D^*\bar D_1$ to $[D^*\bar D]_{L=1}$.
The dotted curve shows the effect of doubling the exponent of the
form factor.
(b) Fit to Belle data  with a resonance; the
dashed curve shows the predicted peak in $D^*\bar D_1$, but scaled
upwards for visibility}
\end{center}
\end{figure}
Fig. 5(a) show the result for  $\rho(s)$ and its dispersion integral
$g^2j(s)$ with a subtraction at  3.95 GeV, well below the
$D^*\bar D_1$ threshold.
The full curves of Fig. 5 show $\rho (s)$ (normalised to
1 at its peak) and its dispersion integral $g^2j(s)$ (normalised
accordingly).
The dashed curve shows the result of a large change in
$F^2$ to $\exp (-6k^2_2)$.

The Belle data show a $Z(4430)$ peak accurately coincident with
the $\bar D D^*$ threshold.
The data can easily be reproduced as a resonance,
and the fit is shown in Fig. 5(b).
Close \cite {Close2} argues in favour of binding by $\pi $ exchange,
though Liu et al. reach the opposite conclusion \cite {Liu}.
Li, L\"u and Wang raise the interesting possibility of a
nonet of strange relatives at the $D_s^*(2112)\bar D_1(2420)$ and
$D^*\bar D_{s1}$ thresholds near 4540 MeV and the $D^*D_{s1}$ threshold
at 4647 MeV.

The width of $D_1(2420)$ reduces the sharpness of the cusp,
though this could be more than compensated by the large number
of decay channels into which $D^*\bar D_1$ may de-excite.
A cusp plus meson exchange could generate an exotic resonance,
but the net attraction must overcome zero-point energy.
We proceed to test whether an alternative fit is
possible using a bare cusp.

\subsection{A subtlety in the equations}
At this point, a digression is needed to
discuss a missing element in T\" ornqvist's equations.
For a 2-channel system (which is sufficient to reveal the
essential point), the usual expression relating the $T$-matrix
and $K$-matrix elements is:
%Eqs. 24 and 25
\begin {eqnarray}
T\rho &=& \frac {1}{B}\left(
\begin {array} {cc}
k_{11} & k_{12} \\
k_{21} & k_{22}
\end {array}
\right)
\left(
\begin {array} {cc}
(1 - ik_{22}) & -ik_{12} \\
-ik_{21} & (1 - ik_{11}
\end {array}
\right)                  \\
B &=& 1 - ik_{11} - ik_{22} + (k_{12}k_{21} - k_{11}k_{22}),
\end {eqnarray}
where $k_{ij}=\sqrt {k_ik_j}a_{ij}$.
Using time-reversal invariance, $k_{12} = k_{21}$. Then
%Eqs. 26 and 27
\begin {eqnarray}
T_{22}\rho _2 &=& \frac
{a_{22}\rho_2 - i\rho_1\rho_2 (a^2_{12}-a_{11}a_{22})}
{1 - ia_{11}\rho_1 - ia_{22}\rho_2 +
\rho_1\rho_2 (a^2_{12}-a_{11}a_{22})} \\
T_{12}\sqrt {\rho_1 \rho_2}&=& \frac {a_{12}\sqrt {\rho_1 \rho_2}}
{{1 - ia_{11}\rho_1 - ia_{22}\rho_2 +
\rho_1\rho_2 (a^2_{12}-a_{11}a_{22})}},
\end {eqnarray}
$T_{11}$ is like $T_{22}$ with indices interchanged.
In order to arrive at a form $T = N/D$ where $N$ is real, it is
necessary to multiply top and bottom of $T_{22}$ and $T_{11}$ by the
complex conjugate of the numerator.
Note that the resulting denominator $D$ is then different from that of
$T_{12}$.
So the result does not describe a resonance, for which $D$ should
be common to all channels.
For all denominators to be the same, the requirement is that
$a^2_{12}-a_{11}a_{22}=0$.
This is satisfied at a pole.
This term was omitted by T\" ornqvist and from the equations of Section
2 because there are indeed second sheet poles very close to the
experimental peaks.
In general however, this term is not zero.
If it is completely dominant, $T_{11}\rho_1$ and $T_{22}\rho _2 \to i$.
In particular, this arises if $T_{12}$ is large.
That is to be expected for de-excitation of $D^* \bar D_1$ to many
$D\bar D$ channels with lower thresholds.

Fig. 6(b) shows the fit to Belle data with a bare cusp and Fig. 6(a)
shows the Argand diagram for $T_{12}$.
Below threshold, the cusp effect creates attraction,
pulling all amplitudes around the periphery of the unitary circle
towards the threshold.
At the threshold, the amplitude curls
towards the centre of the Argand diagram.
As the inelasticity grows, amplitudes for individual decay channels
move back to the origin, though their total goes to $i$.
For this fit, there is no second or third-sheet pole anywhere in the
vicinity of the cusp.
%Fig. 6
\begin{figure}[htb]
\begin{center}
\vskip -12mm
\epsfig{file=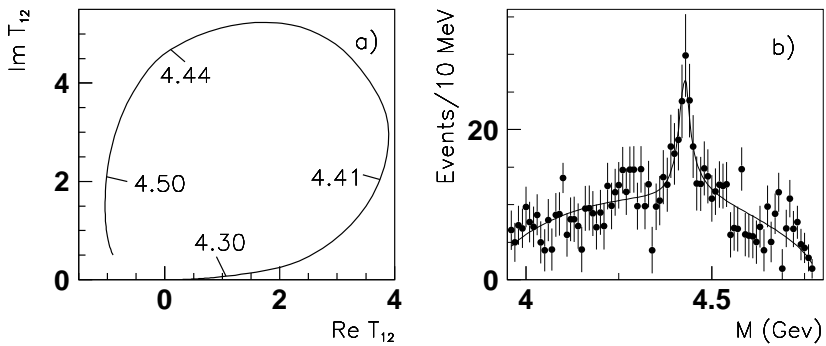,width=12cm}
\vskip -7mm
\caption {Fit with a bare cusp: (a) Argand diagram, (b) fit
to Belle data}
\end{center}
\end{figure}

Some further detail is needed on exactly how data are fitted.
The tails of a cusp depart somewhat from a resonance form.
To allow for this, it is necessary to make small adjustments to
the background fitted by Belle.
This background is taken to be linear in $s$ (slightly higher
at low $s$ than high $s$), multiplied by the phase space factors
$Q_1$ for $Z \to \psi ' \pi$ and $Q_2$ for $B \to K Z$.
The calculation of $a_{ij}$ is done for $Z \to [\bar D D^*]_{L=1}$
as a representative example.
The dispersion integral is strictly over $\rho _1 \rho _2$, but
the dependence on $\rho_1$ is in practice small.
It is assumed that $a_{12} \gg a_{22}$ though the coupling constant
to $\bar D D^*$ is fitted explicitly and so is $a_{11}$, which is
taken as a constant since it is insensitive to form factors.
The denominator of $T_{12}$ for the $\Psi '\pi$ channel is the same
as for all other channels.

How can experiment resolve a cusp from a resonance?
The difference in line-shape is small and easily confused with
experimental backgrounds.
The essential feature of a resonance is that the phase shift
increases by $90^\circ$ up to the centre of the cusp.
The detailed $s$-dependence on Fig. 6(a) is different from a
resonance, but the difference is delicate.
This can in principle be measured from interference with other
components in the Dalitz plot, but would require a
large increase in statistics.

The $Z(4430)$ may be a resonance.
The Argand diagram of Fig. 6(a) resembles a resonance.
If meson exchanges are attractive, it is quite possible that they
will combine with the cusp mechanism to generate a resonance.
Mixing with a genuine diquark-antidiquark system is less likely,
since the long-range tail of the wave function must be purely
molecular.

\subsection {Other points}
Rosner and also Meng and Chao remark on the experimental fact that no
signal is seen for $Z(4430) \to J/\Psi \pi$ despite larger phase space.
This  may be due to a cancellation in the matrix element for
$D^*\bar D_1 \to J/\Psi \pi$.
The momenta for production of $J/\Psi$ and $\Psi'$ are
large: 1130 and 670 MeV/c respectively.
The $J/\Psi $ and $\Psi'$ wave functions are multiplied by a factor
$ \exp (ikr)$ which has nodes at 0.27 or 0.43 fm in the two cases.

In more detail, the matrix element for $Z \to \Psi ' \pi$ in the rest
frame of the $\Psi '$ is
%Eq. 28
\begin {equation}
M = \int ^\infty _0 r^2 dr \, \Psi '(r) [j_1(kr) \cos \delta +
n_1(kr) \sin \delta ],
\end {equation}
using for the initial state the long-range part of its wave
function, expressed in terms of spherical Bessel and Neumann functions.
This is only a first approximation and the short range part of
the wave function is unknown.
The appearance of $n_1\sin \delta$ signifies the fact that
the radial wave function is sucked in by the attractive interaction
due to the cusp, causing the phase shift $\delta$.
For a strong cusp due to many open channels, $n_1 \sin \delta$ should
dominate.
For the kinematics of $J/\Psi $ production, $n_1$ has a node
at 0.38 fm, leading to a distinct cancellation within the matrix
element.
For $\Psi'$ production, $n_1$ has a node at 0.64 fm. This gives
a reasonable overlap with the expected wave function for $\Psi'$,
though by no means perfect.
This discussion is only semi-quantitative because
actual radial wave functions are unknown, but provides
some explanation why the decay to $J/\Psi \pi$ may be weak.

The next task for experiment is a spin analysis of $Z(4430)$, so
expressions will be given for partial waves.
It is highly desirable to use the amplitude for the full
process $B \to K\Psi ' \pi$; both production and decay in principle
determine $J^P$, but angular correlations between them provide
further delicate information which constrains the analysis strongly.
My experience is that 100 fully analysed events achieve the same
significance level (in terms of $\chi^2$ of the fit to data)
as 400--800 events where only information from decays is used.

Formulae simplify greatly in the rest frame of the $Z(4430)$.
There, the angular momentum of the kaon in the production
reaction is expressed by its 3-momentum $\vec K$.
Likewise, the angular momentum of the pion in the decay
$Z \to \psi '\pi$ is given by its 3-momentum $\Pi$.
However, a small correction is required for the Lorentz transformation
of $\Psi '$ spin between its rest frame and that of $Z(4430)$.
Suppose the $xz$ plane is defined by the recoil pion and leptons from
the decay of $\Psi '$.
Let $R$ be the angle between the (unmeasured) polarisation vector and
this plane.
In this plane, let $\theta$ be the angle between the lepton axis and
the pion in the rest frame of $\Psi '$.
The polarisation vector $e$ of the $\Psi ' $ is orthogonal to the
lepton axis in the $\Psi '$ rest frame and may be written as a 4-vector
$e = [\cos R, \sin R, 0, 0]$.
The intensity is obtained using $<\cos ^2 R> = <\sin ^2 R> = 1/2.$
A simple trick avoids this integration.
The $e_x$ and $e_y$ components may be replaced by $1/\sqrt {2}$ and
$i/\sqrt {2}$.
Intensities are then obtained by taking the modulus squared of matrix
elements.
Appendix 1 of \cite {Bugg5} shows that the Lorentz boost between the
rest frames of $\Psi '$ and $Z(4430)$ gives a polarisation vector
$e' = (1/\sqrt {2})[1 + (\gamma - 1)\sin ^2\theta, i,
\sin \theta \cos \theta (1 - \gamma) i,
\beta \gamma \sin \theta \cos \theta ]$,
where $\beta$ and $\gamma$ are the usual parameters of the Lorentz
transformation.
In the $Z$ rest frame, the fourth component of $e'$ drops out of all
matrix elements.
Numerically, $(\gamma - 1) = 0.064$, so the effect of the Lorentz boost
is small.

After this preliminary, expressions for matrix elements $M$ in the
$Z$ rest-frame are simple.
For $J^P = 0^-$, $M = e'.\Pi$ and $K$ does not contribute.
The choice of $xz$ plane gives $\Pi _y = 0$, so the angular
distribution is
$d\sigma /d\Omega \propto (e'_x \Pi_x)^2 + (e' _z\Pi _z)^2
\simeq \Pi ^2_x$.
If the small term in $(\gamma - 1)^2$ is dropped,
this is proportional to $\sin ^2\theta$, as Rosner remarks.

For $J^P = 1^-$, $M \propto K.e'\wedge \Pi = -e'.K\wedge \Pi$,
so the kaon and pion are preferentially orthogonal in the
rest frame of $Z$.
For $J^P = 2^-$, $M \propto \tau _{\alpha \beta}T^{\beta \alpha}$,
where $\tau _{\alpha \beta} = K_\alpha K_\beta - (1/3)
(K_x^2 + K_y^2 + K_z^2)$, representing $L = 2$ for the kaon;
$T_{\alpha \beta} = e'_\alpha \Pi_\beta + e'_\beta \Pi_\alpha
- (2/3)(e'_x\Pi_x + e'_z\Pi_z)$, remembering that $\Pi _y  = 0$.
If only one spin is present, $\sim 100$ events may well be sufficient
to identify the spin.
More than one spin would suggest a non-resonant cusp.

\section {Other cases}
The association of some peaks with thresholds may be numerical
accidents.
Hence examples discussed here are not comprehensive, but
concentrate on cases where the agreement is remarkably close or is
otherwise of special interest.

A narrow peak is observed in $D^0p$ by Babar \cite {B2940} and Belle
\cite {Bel2940} with a mass of $2939.8 \pm 1.3 \pm 1.0$ MeV and
$\Gamma = 17.5 \pm 5.2 \pm 5.9$ MeV and also in $\Sigma _C(2455)\pi$.
Its mass is just below the $D^0p$ threshold at $2944.9 \pm 0.4$ MeV.
For an S-wave threshold, $J^P = (1/2)^-$ or $3/2^-$.
It could be a molecular state \cite {HeLiu}; alternatively a nearby
regular $\Lambda_C$ state may be captured by the $D^0p$ threshold.

The $\Lambda _C^+(2595)$ has a mass of $2595.4 \pm 0.6$ MeV \cite {PDG},
very close to that of $\Sigma_C^{++}\pi ^-$, namely 2593.6 MeV.
It is too narrow to be a pure cusp, but may be a resonance attracted to
the threshold where decays to $\Lambda _C\pi$ are possible.

The $\Psi (4039)$ lies close to the $D^*\bar D^*$ threshold at
4014 MeV.
However, it has a width of $80 \pm 10$ MeV, making its association
with a cusp effect questionable.
Dunynskiy and Voloshin discuss in detail the complicated
dependence on mass of $D^*\bar D^*$, $D^0 \bar D^0$ and $D_s \bar D_s$
channels \cite {Voloshin}.

The $D_s(2315)$ is very narrow and $\sim 50$ MeV below the
$D\bar K$ threshold.
It is therefore unlikely to be a molecular state, though it could
be an example of a diquark-antidiquark configuration.
In the absence of other clear examples of strongly bound 4-quark
systems, the likely explanation is a $c\bar s$ state.

Amongst the light mesons, the $f_2(1565)$ has a phase variation
which is well determined by Crystal Barrel data on $\bar pp \to
3\pi ^0$ at rest.
Those data have been fitted simulaneously with data on $\bar pp \to
\omega \omega \pi ^0$ at rest \cite {Baker} and definitely require
a resonance accurately at the $\omega \omega $ threshold.
The $f_2(1640)$ reported by the Particle Data Group is the
decay to $\omega \omega$, which is moved upwards from 1565 MeV
by $\omega \omega$ phase space.
There is no other candidate for the radial excitation of $f_2(1565)$,
so it appears that this state has been captured by the
$\omega \omega$ threshold.
Note, however, that the well known $f_0(1500)$ is not attracted to
that threshold; its decays into $\omega \omega$ are weak.

The $\rho (1900)$ of the Particle Data Tables has a rather narrow
width.
This is suggestive of a cusp due to the strong $\bar pp$ threshold.

The $\pi _1(1400)$ could be a bare cusp.
It has been assumed by most groups to be a resonance.
However, it is close to the $f_1(1285)\pi$ and $b_1(1235)\pi$
thresholds, which would appear in S-wave decays.
Dzierba et al. question whether it is a threshold effect or a
resonance \cite {Dzierba}.
The $\pi_1(1600)$ appears at higher mass with the same quantum
numbers and is observed dominantly in $b_1(1235)\pi$ and less
strongly in $f_1(1285)\pi$.
So the $\pi_1(1400)$ could be a molecular configuration coupled to
these thresholds or could be simply a cusp effect.
A full analysis is needed of the two alternatives using analytic
forms like those presented here.
The data on $\bar pp \to \pi ^0 \pi ^0\eta$ show only a very
weak signal in the $\eta \pi$ P-wave, insufficient to tell the
difference between the two alternatives \cite {BASZ}.

Valcarce, Vijande and Barnea have made an interesting study
of mixing between diquark and tetraquark configurations
\cite {Vijande}, though they do not specifically take the cusp
effect into account.

\subsection {$X(1812)$}
An intriguing case is the sharp $\omega \phi$ signal reported by
the BES collaboration \cite {omegaphi}, peaking at 1812 MeV, just
above the $\phi \omega$ threshold at 1801 MeV.
If it were purely a threshold effect, it should peak considerably
higher.
It is therefore very likely associated with the $f_0(1790)$,
a resonance clearly separated from $f_0(1710)$ in BES data on
$J/\Psi \to \omega KK$ \cite {WPP} and $\phi \pi \pi$ \cite {phipp}.
Data on the $\omega KK$ channel display a strong $f_0(1710)$ peak,
but nothing is visible in $\omega \pi \pi$, despite large statistics.
Conversely, the $f_0(1790)$ appears clearly in $\phi \pi \pi$,
but any $\phi KK$ signal is weak.
There is a factor 22 difference in decay branching ratios to
$\pi \pi$ and $KK$, hence requiring separate $f_0(1710)$ and
$f_0(1790)$.
The $f_0(1790)$ is also observed in $J/\Psi \to \gamma 4\pi$ \cite
{Mark3} \cite {g4pi}.
It is accomodated naturally as the radial excitation of $f_0(1370)$.

The $\phi \omega$ decay can arise naturally from a glueball component
in $f_0(1790)$ \cite {Bicudo} \cite {BuggG}.
A glueball is a flavour singlet.
It has flavour content
%Eq. 29
\begin {equation}
F = (u\bar u + d\bar d + s\bar s)(u\bar u + d\bar d + s\bar s).
\end {equation}
If the decay is to vector mesons, the component $(u\bar u + d\bar d)
(u\bar u + d\bar d)$ makes three charge combinations of $\rho \rho$
and one of $\omega \omega$.
The component $2(u\bar u + d\bar d)s\bar s$ can make $4\omega \phi$
or $2(K^{*0}\bar K^{*0} + K^{*+}K^{*-})$ or some linear combination.

There are BES I data on $J/\Psi \to \gamma
(K^+\pi ^- K^-\pi ^+)$  showing that the channel  $\gamma (K^*\bar K^*)$
does not contain any significant $0^+$ signal \cite {KstKst}.
The paper says: `Contributions from $0^{++}$ and $4^{++}$ are small
or absent'.
A signal with the same magnitude as that of
$J/\Psi \to \gamma (\omega \phi)$ in \cite {omegaphi} would be rather
conspicuous near 1800 MeV, because of the small phase space at that
mass for $K^*\bar K^*$.
Its absence there may be qualitatively attributed to the fact that
$(u\bar u + d\bar d)s\bar s$ has larger
phase space in $KK$ decays than $K^* \bar K^*$.

\section {Concluding Remarks}
A sharp threshold generates a cusp in the real part of scattering
amplitudes at the opening of a new threshold.
This is a dispersive effect, arising from analyticity.
If there is an attractive $t$- or $u$-channel exchange, it can
add coherently to the cusp effect and generate a resonance.
The $f_0(980)$ appears to behave in this way.
The cusp can also add coherently to the confinement `potential'
and attract a regular quark resonance to the threshold.
This explains why states like $f_2(1565)$, $K_0(1430)$ and
$\Lambda _C(2940)$ appear at thresholds.

At the threshold, zero point energy is minimised by the long-range
tail of the wave function.
Mixing between quark configurations and meson-meson states minimises
the energy of the linear combination in a way analogous to the
formation of a covalent bond in chemistry.

Diquark-antidiquark resonances may exist.
However, if they lie close to a threshold, zero-point energy
will necessarily mix a large molecular component into the wave
function.
The $f_0(980)$, for example, has a component of $\sim 60\%$ KK.
Section 2 used known parameters of $f_0(980)$ to examine how the
second-sheet pole position is affected by perturbations to its
parameters.
The conclusion is that the cusp mechanism can attract a resonance
over a surprisingly large mass interval of order $\pm 100$ MeV.
If the open channel ($\pi \pi )$ is switched off, the resonance
becomes a bound-state pole just below threshold.

The $X(3872)$ is too narrow to be fitted as a pure cusp.
Although parameters of the cusp can be fine-tuned to fit the
line-shape observed in decays to $\rho J/\Psi$, they then fail
to reproduce the peak observed 3.5 MeV higher in $\bar D D^*$ decays.
Both peaks may be fitted with a resonance or virtual state.
A definitive distinction between these possibilities requires phase
information from interferences in Dalitz plots: for a virtual state,
the phase falls above the threshold.
The natural explanation of $X(3872)$ is that the $\bar ss$ $^3P_1$
radial excitation has been attracted to the $\bar D D^*$ threshold.
The weak decays of $c\bar c$ to non $\bar DD$ channels like
$\rho J/\Psi$, $\omega J/\Psi$, $[\chi \sigma]_{L=0}$ and
$[\eta _C \sigma ]_{L=1}$ lead to a very narrow resonance.

The $Z(4430)$ can be fitted as a resonance.
It is quite possible that meson exchanges generate sufficient
attraction to turn the cusp into a resonances.
However, it cannot presently be excluded  that a non-resonant cusp
fits the data using equations (26) and (27).
The availability of many de-excitation channels such as $[\bar DD^*
]_{L=1}$ necessarily produces a strong cusp.
Partial wave formulae to assist separation of spin-parity assignments
$0^-$, $1^+$ and $2^-$ are given in Section 4.
The presence of more than one $J^P$ would suggest a bare cusp.

Equations (8), (9) and (14) for a sharp threshold improve on the
Flatt\' e formula and are just as easy to use.
The experimental data for $f_0(980)$ conform with the line-shape
predicted by these equations.
However, as T\" ornqvist remarks, increasing the precision of the
formula and its parameters may be an academic exercise.
A case which does require study with full inclusion of the
dispersive effect is $\pi _1(1405)$, which could be a resonance or
could be just a threshold cusp.

\section {Acknowledgements}
I wish to thank Prof. G. Rupp and Prof. E. van Beveren for
extensive discussions.
Use of their model led to important insight into the way poles
move near a threshold.
It is recommended as a test-bed for quantitative studies.
I also wish to thank Dr. Y. Kalashnikova for illuminating
comments on the Flatt\' e formula.

\end{document}